\documentclass[12pt,preprint]{aastex}
\usepackage{emulateapj5}
\usepackage{apjfonts}
\usepackage{epsfig}

\submitted{ApJ, accepted}

\def\ltsima{$\; \buildrel < \over \sim \;$}
\def\simlt{\lower.5ex\hbox{\ltsima}} 
\def\gtsima{$\; \buildrel > \over \sim \;$}
\def\simgt{\lower.5ex\hbox{\gtsima}} 

\shorttitle{NIR Study of BL Lac Host Galaxies}
\shortauthors{Cheung et al.}

\def\hst{{\it HST}} 

\begin{document}

\title{Near-Infrared Observations of BL Lacertae Host Galaxies}

\author{C. C. Cheung\altaffilmark{1,4}, C. Megan Urry\altaffilmark{2,4},
Riccardo Scarpa\altaffilmark{3,4} \& Mauro Giavalisco\altaffilmark{4}}

\altaffiltext{1}{Department of Physics, MS\,057,  
Brandeis University, Waltham, MA 02454; ccheung@brandeis.edu}
\altaffiltext{2}{Yale Center for Astronomy \&
Astrophysics, Department of Physics (JWG 460), Yale University, New Haven,
CT 06520; meg.urry@yale.edu}
\altaffiltext{3}{European Southern Observatory, Avenida  
Alonso de Cordova 3107, Vitacura, Casilla, 19001, Santiago 19, 
Chile; rscarpa@eso.org}
\altaffiltext{4}{Space Telescope Science Institute, 3700 San Martin Drive,
Baltimore, MD 21218}

\begin{abstract}

Multi-band near-infrared images of twelve BL Lacertae objects were
obtained with the 2.5m telescope at the Las Campanas Observatory in order
to determine the properties of their underlying host galaxies.  Resolved
emission was clearly detected in eight of the lowest redshift targets (up
to $z\sim$0.3), and was modeled with a de Vaucouleurs $r^{1/4}$ surface
brightness law.  We find that the morphologies match the elliptical galaxy
profiles well, and that the BL Lac objects reside in large and luminous,
but otherwise normal hosts -- consistent with previous studies done
predominantly at optical wavelengths.  The median absolute K-band
magnitude of the galaxies in this study is --26.2, the average half-light
radius is $4.2 \pm 2.3$ kpc, and their average integrated $R-K$ color is
$2.7 \pm 0.3$ mag.  These are well within the range of values measured
previously in the H-band by Kotilainen et al. and Scarpa et al. in a
comparable number of targets.  Taking their data together with our
results, we find a best-fit K-band Kormendy relation of $\mu_{e} =$ 4.3
log$_{10}$ ($r_{e}$/kpc) $+ 14.2$ mag arcsec$^{-2}$, virtually identical 
to that
obtained for normal ellipticals.  Finally, the near-infrared colors
determined for five galaxies (average $J-K$ = $0.8 \pm 0.3$ mag) are the
first such measurements for BL Lac hosts, and match those expected from
old stellar populations at the BL Lac redshifts.

\end{abstract}


\section{Introduction}

The study of the host galaxies of BL Lacertae objects in recent years has
contributed much to our understanding of these enigmatic objects.  
According to unified schemes for active galactic nuclei (AGN), BL Lacs are
low-luminosity \citet{Fan74} type I (FRI)  radio galaxies whose jets are
aimed very close to our line of sight \citep[e.g.,][]{Urr95}.  This
explains their observed extreme properties primarily by relativistic
beaming in the jet.  The host galaxies and environments of BL Lacs should
therefore be independent of orientation effects and should be similar to
those of the hosts of FRIs.  Indeed, observational data confirm that BL
Lac objects are predominantly found at the center of giant, but otherwise
normal, elliptical galaxies \citep[e.g.,][]{Wur96,Urr00,Sca00a}.  Also,
the available integrated colors and color profiles in a number of these
galaxies reveal stellar populations typical of normal ellipticals at the
appropriate redshifts \citep{Kot98b,Urr99,Sca00b}, consistent with the
bulk of the star formation having occurred at the redshifts of at least a
few.

Here, we report on moderately deep near-infrared (NIR) observations of
twelve BL Lacs drawn from the larger sample of 110 objects observed
(mostly in R-band) in the {\it Hubble Space Telescope} (\hst ) WFPC2
survey of BL Lac objects \citep{Sca00a,Urr00,Fal00}. The survey targets
were drawn from seven complete radio, optical, and X-ray flux limited
samples. A subset of the sample was observed with NICMOS at H-band
\citep{Sca00b}, and there are two objects in common with the present
observations (MS\,2143+070 and H\,2356--309). Our relatively long
ground-based NIR observations had the advantage over the NICMOS data that
we were able to follow the surface brightness profiles of each object out
to larger radii. NIR filters give us greater sensitivity to older, and
therefore redder stellar populations compared to optical filters.  The
number of targets in this study is comparable to the H-band samples
studied by \citet{Kot98b} and \citet{Sca00b}, and we utilized their
results for comparison.  Following much of the previous literature on this
subject, we use $H_{0}$ = 50 km s$^{-1}$ Mpc$^{-1}$ and $q_{0}$ = 0
throughout this paper.

\section{Observations and Data Reduction}
\label{section:observations}

Twelve BL Lac objects were observed with the du Pont 2.5m f/7.5 telescope
at the Las Campanas Observatory on four nights from 28 September to 1
October 1998 using the 256$\times$256 sq. pixel NIR camera \citep{Per92}
at medium resolution, with a scale of 0$\arcsec$.348/pixel.  The seeing
during our observations was typically below 1$\arcsec$ FWHM
(Table~\ref{tbl-1}).  We used the J$_{s}$ (1.24 $\mu$m), H (1.65 $\mu$m),
and K$_{s}$ (2.16 $\mu$m) filters, where J$_{s}$ and K$_{s}$ are variants
of the respective standard bands designed to reduce the background signal
at these wavebands \citep{Per98}, and H is the H-band filter of the
CIT/CTIO photometric system.  Differences between the J$_{s}$- and
K$_{s}$-bands from the standard J and K are negligible and we refer to
only the standard bands throughout the rest of this paper.  We observed
two BL Lacs in all three NIR bands, three were observed in both J and K,
and the remaining seven were imaged in K only.  The observations are
summarized in Table~\ref{tbl-1}.


We used the NOAO Image Reduction and Analysis Facility (IRAF\footnote{IRAF
is distributed by the National Optical Astronomy Observatories, which are
operated by the Association of Universities for Research in Astronomy,
Inc., under cooperative agreement with the National Science Foundation.})  
package to calibrate the data. Raw exposures were dark subtracted and flat
fielded. Flats for each filter were made by median combining several night
sky exposures -- usually referred to as ``super sky flats''. Bad pixels
were identified via a mask image made from a ratio of two real sky flats
with different exposure times using the task \verb+CCDMASK+. These bad
pixels were substituted with interpolated values using the task
\verb+FIXPIX+.

Due to the high and widely varying sky brightness at NIR wavelengths, we
dithered our exposures over nine positions in a square grid on the CCD to
construct the sky frames, usually acquiring several short exposures at
each position in order to prevent saturation.  Each dithered image was sky
subtracted, shifted relative to the center image, and then
median-combined. In most cases, a series of these dithered exposures of
the target were taken consecutively in the same night and were combined.  
Observations of standard stars taken from \citet{Per98} were used to set
the photometry.  We found an inconsistency in the photometry for one star
(SS\,9175) used during the last night of our observing run. The chart for
this object is suspected to have been printed incorrectly (E. Persson,
2000, private communication), therefore we simply did not use it in our
calibration.  The RMS in the zero-points derived from the three other
standards within each night was smaller than 0.1 mag in all three bands.  
We found also that the standards were stable to better than 0.1 mag over
the course of the four night observing run; we therefore estimate
conservatively that our 1$\sigma$ error in photometry is 0.1 mag.


In order to maximize the signal from the underlying hosts in our analysis,
we extracted azimuthally averaged one-dimensional surface brightness
profiles from the final BL Lac images using an interactive script
implemented in \verb+SUPERMONGO+.  The profiles were then fitted with a
point spread function (PSF) plus \citet{dev48} $r^{1/4}$ elliptical galaxy
model convolved with the PSF, using a $\chi^{2}$ minimization technique as
outlined in \citet{Sca00a}.  In all cases (with the possible exception of
H\,1914--194) the hosts were known to be elliptical \citep{Sca00a}, so no
attempts were made to fit the resolved emission with exponential disk
profiles.


Stars within our $\sim1.5\arcmin\times1.5\arcmin$ field of view were often
too dim to extract a PSF with accurate wings.  Since only the central part
of the PSF is affected by the seeing, we combined the innermost
($\sim2\arcsec$) part of the field star profile (or an average when
multiple stars were in the field) in each BL Lac image with the outer wing
extracted from a standard star observed in the same band on the same
night. We found that this technique provided us with reliable descriptions
of the PSF out to $\sim8\arcsec$.

\begin{figure*}
\figurenum{1}
\begin{center}
\epsfig{file=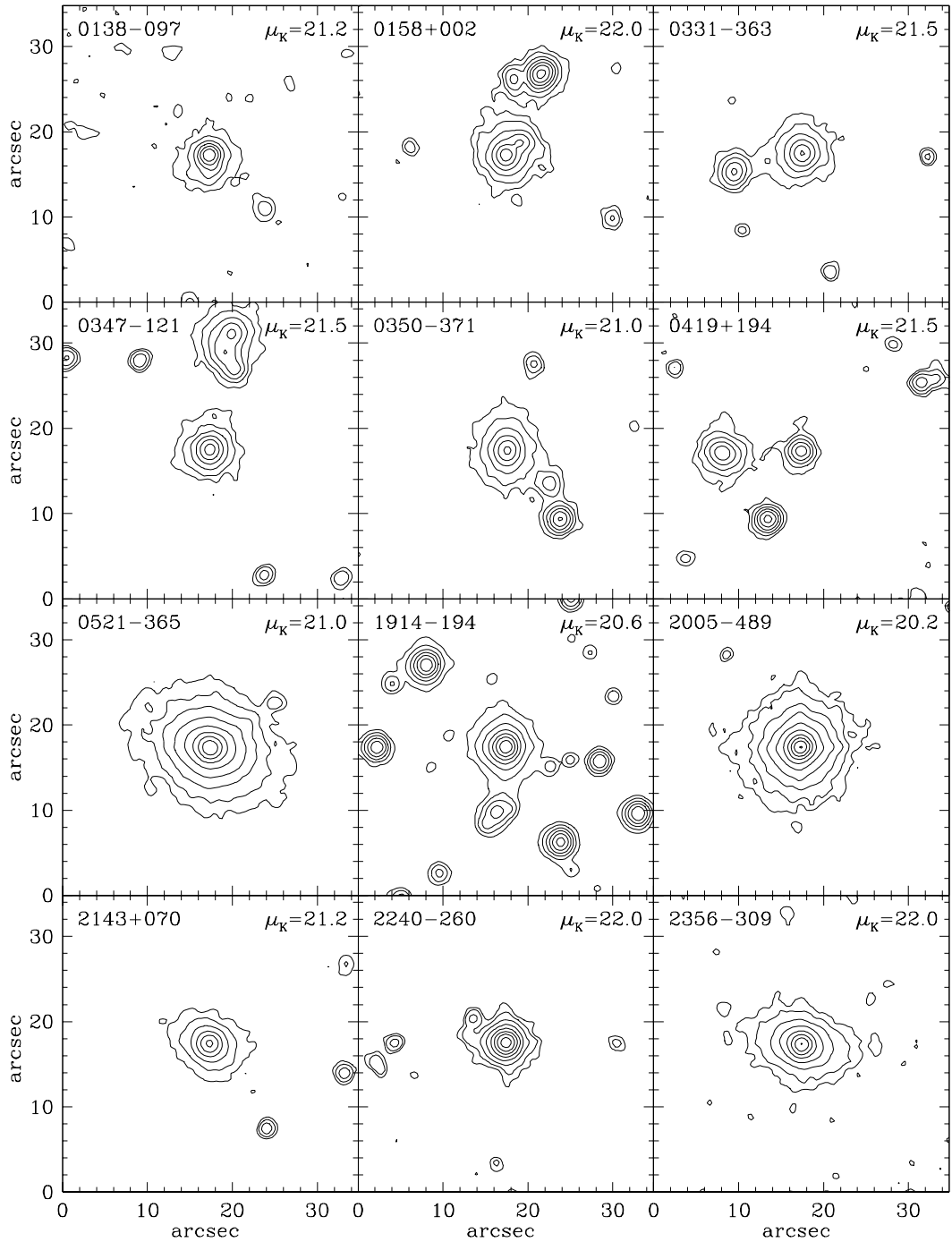,width=5.5in,angle=0}
\end{center}
\figcaption[f1.eps]{\label{fig-1}
K-band contour images of the targets observed after smoothing
with a gaussian filter with $\sigma$= 1 pixel. In each case, a
$\sim35\arcsec\times35\arcsec$ section of the field is shown, centered on
the BL Lac object.  Minimum contour levels (in mag arcsec$^{-2}$) are
indicated on each image and successive isophotes are spaced by one mag.   
North is up on each image. For the objects which we obtained images in two
separate nights, we only show the deeper image obtained with better
seeing.}
\end{figure*}

\begin{figure*}
\vspace{-0.2in}
\figurenum{2}
\begin{center}
\epsfig{file=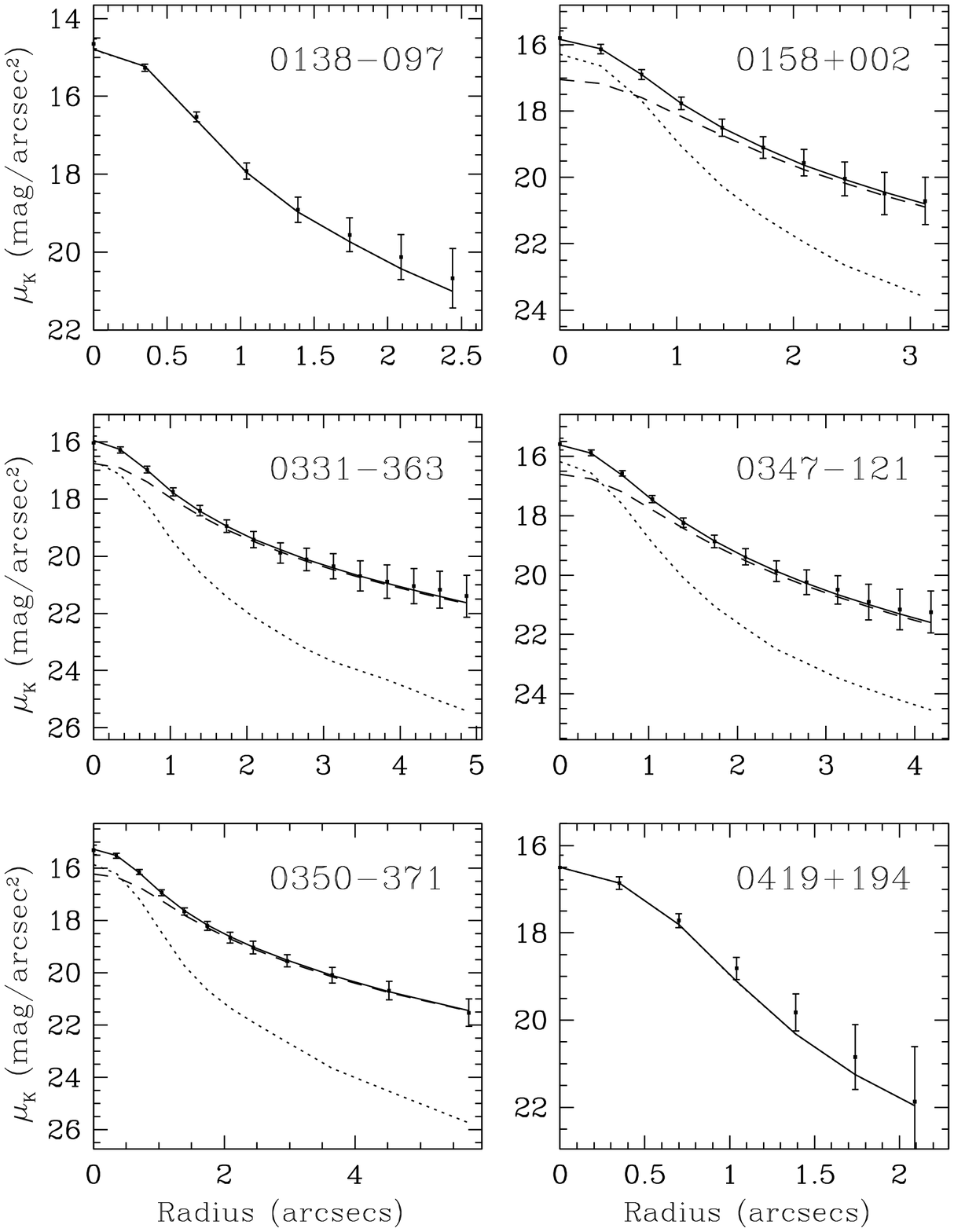,width=3.5in,angle=0}
\epsfig{file=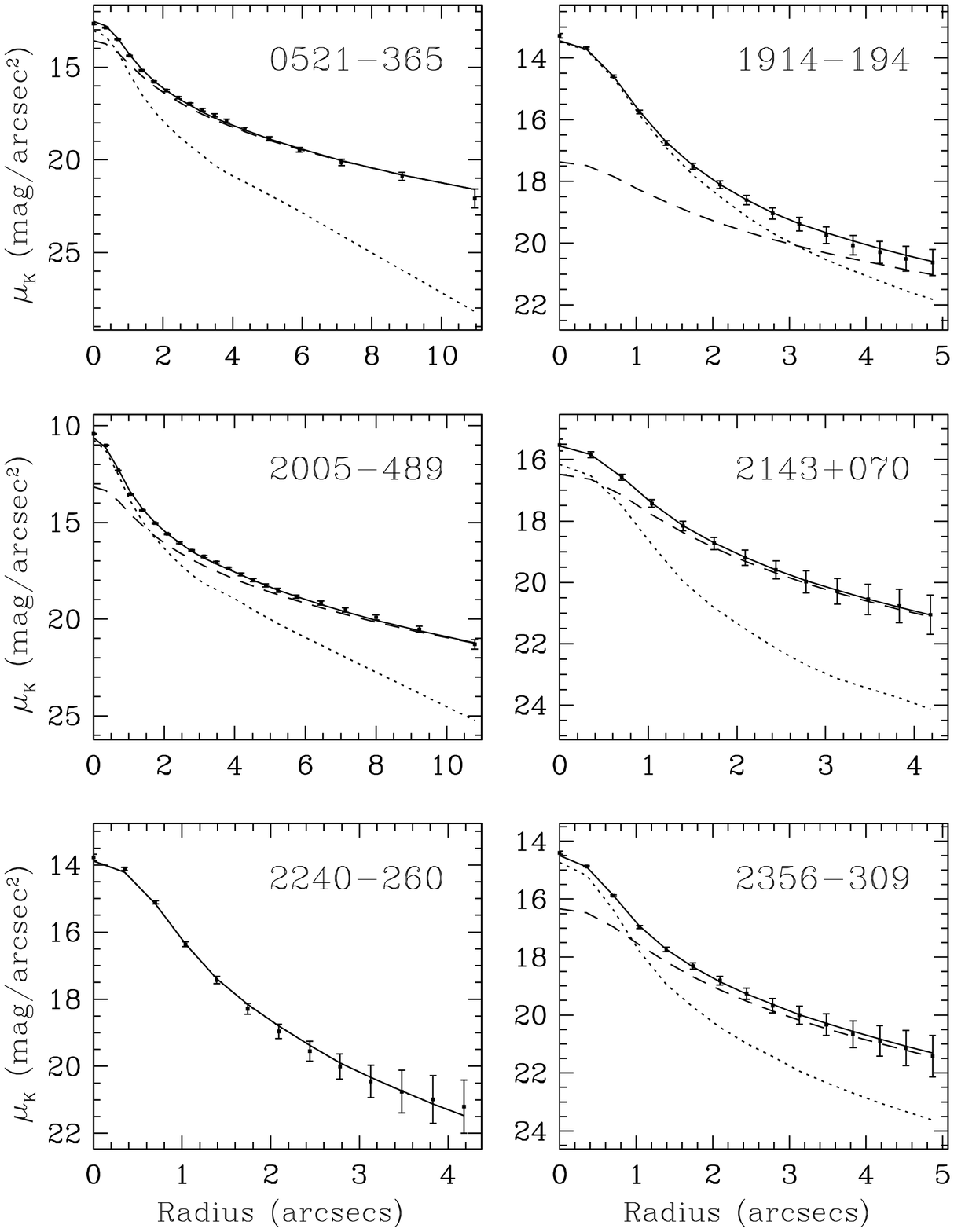,width=3.5in,angle=0}
\end{center}
\figcaption[f2a.eps,f2b.eps]{\label{fig-2}
K-band azimuthally averaged radial surface brightness profiles of the
sources (shown with error bars). These profiles were extracted from the
images presented in Figure~\ref{fig-1}.  Also shown are the scaled PSF
(dotted line), the de Vaucouleurs $r^{1/4}$ galaxy profile convolved with
the PSF (dashed line), and their sum (solid line). Wherever only the solid
line is shown, the host galaxy was unresolved.
}
\end{figure*}

Final K-band images of the twelve objects observed are shown in
Figure~\ref{fig-1}; the available J- and H-band images display very
similar morphology to the K-band images so are not presented. Azimuthally
averaged K-band profiles for the targets extracted from the images in
Figure~\ref{fig-1} are presented in Figure~\ref{fig-2}, and the best fit
model parameters for all bands are reported in Table~\ref{tbl-2}. In
several cases, we observed the same object in the same band on two
separate nights during our observing run. We analyzed these data
independently and found that the de Vaucouleurs plus PSF fits on the data
from the different nights agreed well within our 1$\sigma$ limits. 


\section{Results}
\label{section:results}

The host galaxies of the eight BL Lacs at $z \leq 0.308$ were clearly
resolved in our images, as they were also in the \hst\ optical survey
\citep{Sca00a,Urr00}. The three objects at larger ($z > 0.5$) redshifts
were unresolved (PKS\,0138--097, MS\,0419+194 and PKS\,2240--260) so only
some brief notes are presented for them in section~\ref{subsection:notes}.
In the \hst\ survey data, the host of H\,1914--194 was resolved but was
consistent with both elliptical and disk morphologies.  It is also
resolved (though marginally) in the present data, but there is no robust
determination of its redshift in the literature and further discussion of
this object is limited to section~\ref{subsection:notes}. The derived
properties of several host galaxies studied previously in the H-band by
\citet[][MS~2143+070, H~2356--309]{Sca00b} and \citet[][PKS~0521--365,
PKS~2005--489, MS~2143+070]{Kot98b} are in agreement with our data
(section~\ref{subsection:notes}).

\subsection{Morphologies and Luminosities of the Host Galaxies}
\label{subsection:morphology}

The resolved hosts are large, with average half-light radius ($r_{e}$)
$4.2 \pm 2.3$ kpc.  Although the uncertainties on the half-light radii are
large (with the exceptions being the nearby bright galaxies in
PKS~0521--365 and PKS~2005--489), the derived sizes are consistent with
those derived from similar NIR ground-based observations by \citet{Kot98b}
where they found $<r_{e}> = 4.8 \pm 1.9$ kpc (omitting their large value
of $r_{e} = 32.5$ kpc for PKS~2254+074).  On average, our results are also
consistent within the errors with $<r_{e}> = 10 \pm 5$ kpc found in the
H-band NICMOS study by \citet{Sca00b}, although the ground-based data seem
to prefer smaller values. A direct comparison of the $r_{e}$ derived for
the two objects (MS~2143+070, H~2356--309) common to both the ground and
space-based studies show that the individual values are well in agreement
(see section~\ref{subsection:notes}).

Recently, \citet{Bar03} noted that the half-light radii derived from the
\hst\ R-band survey \citep{Sca00a,Urr00} were systematically smaller than
those obtained through ground-based optical studies in their low redshift
sample of eleven BL Lacs. This difference is most pronounced in the
largest galaxies where they suggest a combination of short exposure time
and relatively small field of view of the survey data may have led to an
underestimation of $r_{e}$ values from the \hst. We do not find such
large discrepancies in our comparison of the NIR data above -- our
comparison actually suggests on average that the situation may be slightly
reversed from what is found in the comparison of optical data, so this
issue remains unresolved.

BL Lac host galaxies have been found to be quite luminous in optical
studies \citep{Fal96,Wur96,Urr00}, with luminosities commensurate with
those of the brightest cluster galaxies \citep{Tay96,Wur96}.  Our median
rest-frame galaxy NIR magnitudes, $M_{K} = -26.2$ mag (with a range of
--26.9 to --25.6) and $M_{J} = -25.4$ mag, are about one magnitude
brighter than the characteristic magnitude for nearby non-active
ellipticals \citep[$M^{*}_{K} = -25.0$ mag;][]{Mob93}.  These observations
are consistent with the H-band values found by \citet{Kot98b} and
\citet{Sca00b}, and overall, confirm results from the previous optically
based studies that the hosts of BL Lac objects are large and luminous, but
otherwise normal, elliptical galaxies.

Despite the fact that we were plagued with large uncertainties in our
measurements of the effective radii, we combined our K-band
$\mu_{e}-r_{e}$ data (Table~\ref{tbl-2}) with the H-band data from
\citet{Kot98b} and \citet{Sca00b} to construct a K-band Kormendy relation
for BL Lac host galaxies. The H-band data were converted to K assuming
$H-K=0.22$ \citep{Rec90}. This linear relation is a projection of the
fundamental plane relating the half-light radius to the surface brightness
at that radius \citep{Djo87,Kor89}.  As a result, we obtained a best
linear fit of $\mu_{e} =$ 4.3 log$_{10}$ ($r_{e}$/kpc) $+ 14.2$ mag
arcsec$^{-2}$ to the data (see Figure~\ref{fig-3}). This relation is
virtually identical to that derived from K-band measurements of a sample
of 59 normal elliptical galaxies ($\mu_{e} =$ 4.3 log$_{10}$ ($r_{e}$/kpc)
$+ 14.3$ mag arcsec$^{-2}$, after converting to our adopted cosmology) by
\citet{Pah95}, and is consistent with the relation found previously using
the H-band data only \citep{Sca00b}. Thus, BL Lac hosts and normal
elliptical galaxies appear to be dynamically similar and occupy the same
region of at least this projection of the fundamental plane.  Lastly, we
fit a similar $\mu_{e}-r_{e}$ trend in both slope ($\sim$4.4) and
intercept ($\sim$13.5) with a fairly large scatter using the resolved and
marginally resolved data for a similarly small sample of flat spectrum
radio quasars \citep{Kot98a}. This appears at least qualitatively similar
to our derived relation for BL Lacs, indicating perhaps that the
luminosity of the central source does not strongly influence the
properties of their surrounding host galaxies.

\begin{center}
\figurenum{3}
\plotone{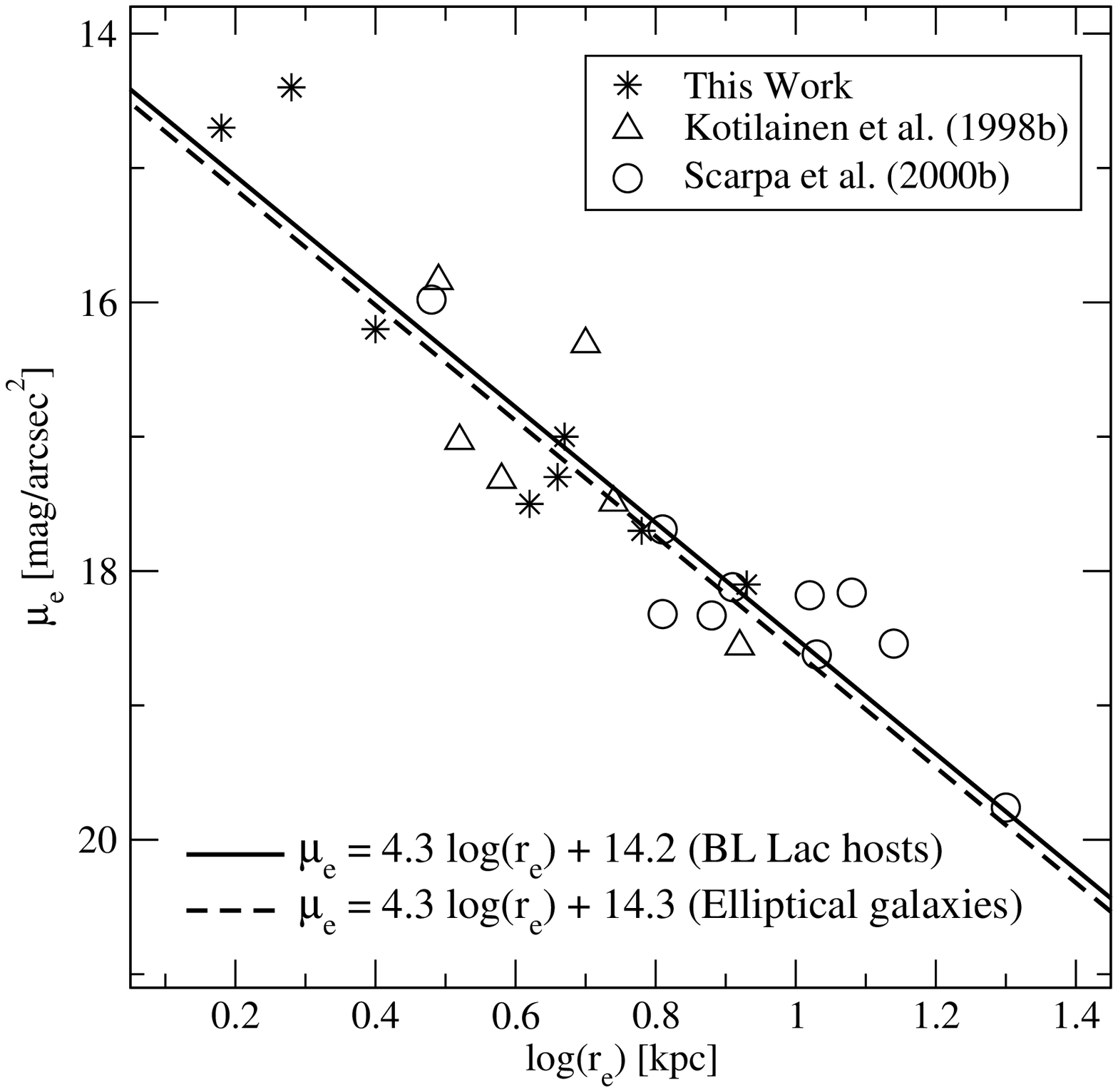}
\vspace{-1.3in}
\figcaption[f3v2.eps]{\label{fig-3}
K-band $\mu_{e}-r_{e}$ Kormendy relation for BL Lac host galaxies using
our K-band results (stars) and the H-band data from
\citet[][triangles]{Kot98b} and \citet[][circles]{Sca00b} which were    
converted to K assuming a $H-K=0.22$ color typical of normal ellipticals  
\citep{Rec90}. The solid line is a linear least-square fit to all of the
data ($\mu_{e} =$ 4.3 log$_{10}$ ($r_{e}$/kpc) $+ 14.2$ mag arcsec$^{-2}$) which 
is virtually identical to that derived for a sample of 59 inactive 
elliptical galaxies in the K-band (dashed line: $\mu_{e} =$ 4.3 log$_{10}$ 
($r_{e}$/kpc) $+ 14.3$ mag arcsec$^{-2}$, after converting to our adopted cosmology) by
\citet{Pah95}, and is consistent with a H-band only relation of $\mu_{e} 
=$3.8 log$_{10}$ ($r_{e}$/kpc) $+ 14.8$ mag arcsec$^{-2}$ obtained previously by
\citet{Sca00b}.}
\vspace{0.1in}
\end{center}

\subsection{Host Galaxy Colors}
\label{subsection:colors}

We calculated integrated rest-frame $R-K$ colors for the 8 of the 9
galaxies resolved in our observations (neglecting H\,1914-194, which has
no measured redshift) utilizing R-band values derived from the uniformly
analyzed \hst\ survey data \citep{Sca00a,Urr00}. The results are reported
in Table~\ref{tbl-3}. Additionally, we have information in two NIR filters
for five of these galaxies, so we were also able to calculate $J-K$
colors. To our knowledge, these are the first NIR colors published for any
BL Lac host galaxy. We obtained average $R-K$ and $J-K$ values of $2.7 \pm
0.3$ and $0.8 \pm 0.3$ mag, respectively. These colors closely match those
found in normal non-active elliptical galaxies:  $<R-K> = 2.7$
mag\footnote{Using $<V-R> = 0.6$ mag \citep{Gre89} and $<V-K> = 3.3$ mag
\citep{Sch93,Bre96}.} and $<J-K> = 0.9$ mag \citep{Sch93,Sil94} and are in
agreement with evolutionary synthesis models \citep[e.g.][]{Pog97}.


Color profiles for the eight resolved galaxies are presented in
Figure~\ref{fig-4}. We used R-band data from the \hst\ survey
\citep{Sca00a,Urr00} and the present K-band results to compute these
profiles. The BL Lac host galaxies show a clear gradient in their colors,
being bluer toward the periphery. This trend was previously found by
\citet{Kot98b} and confirmed by \citet{Sca00b} with additional data for
more objects. We computed an average slope of the profiles to be
$\Delta(R-K)/\Delta log~r = -1.27 \pm 0.73$ (Table~\ref{tbl-3}). Compared
with results from the previous NIR studies, this average color gradient
appears, however, to be much steeper than observed in $R-H$. The magnitude
of the average gradient values is reduced by only about one-half even when
the three galaxies exhibiting extremely large gradients hovering around
--2 are not considered.  The large gradients we computed may be an
artifact due to the quite small effective radii we fit to our NIR data --
our average K-band effective radius is 4.2 kpc, compared with $\sim$10 kpc
in the R-band \citep[e.g.,][]{Fal99,Urr00}, although we can not be certain
of the true cause. For comparison, normal non-active elliptical galaxies
show much less pronounced gradients, but with the same sign, with
typically $\Delta(V-K)/\Delta log~r = -0.16$ \citep{Pel90}, and --0.26
\citep{Sch93}.

\begin{center}
\figurenum{4}
\plotone{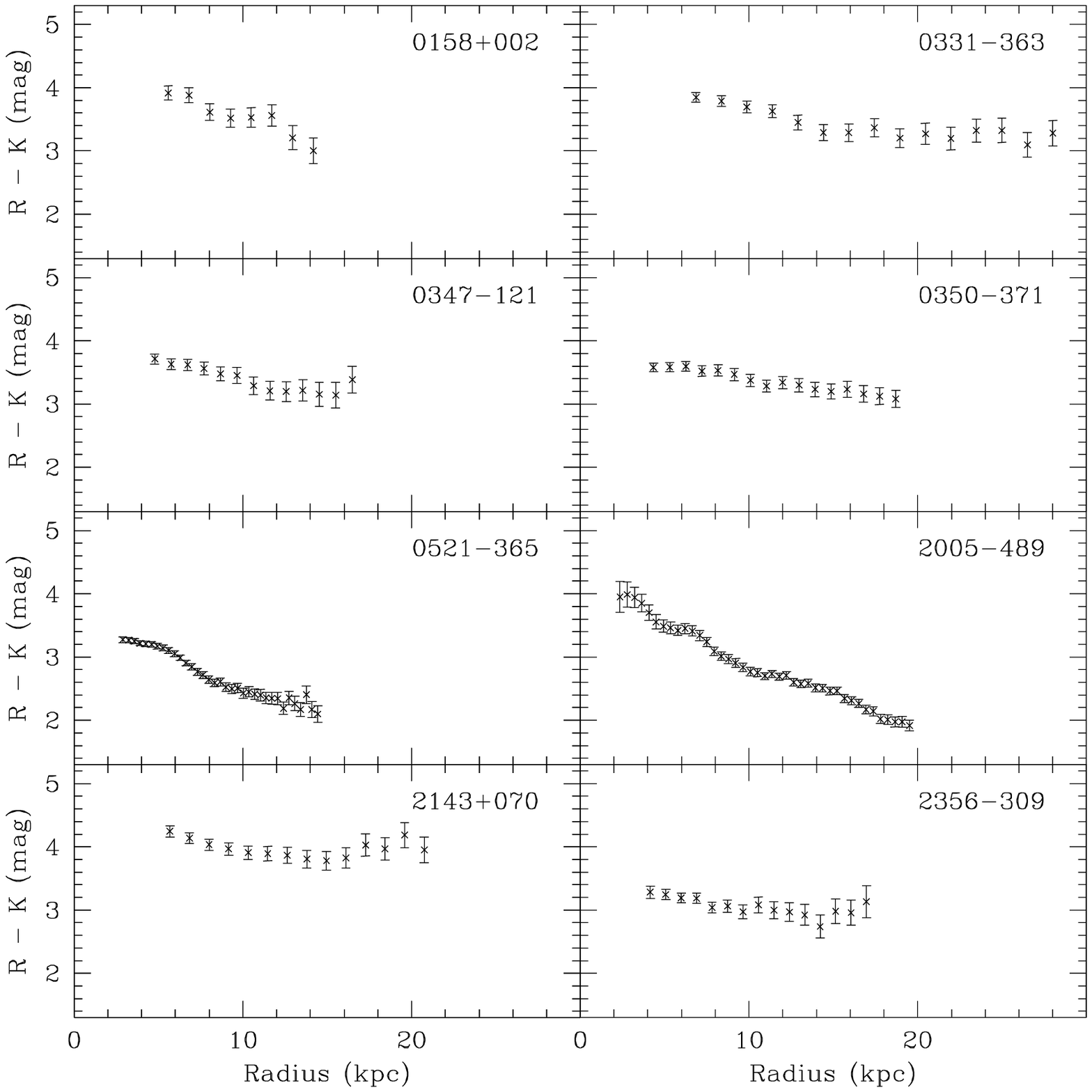}
\figcaption[f4.eps]{\label{fig-4} 
$R-K$ color profiles for the eight resolved host galaxies calculated using
the R-band \hst\ survey data \citep{Urr00,Sca00a} and the K-band data
from the current work. The profiles indicate that the galaxies are bluer
further from the nuclei. We find an average slope of $\Delta(R-K)/\Delta
log~r= -1.27 \pm 0.73$.  See table~\ref{tbl-3} for individual values and 
section~\ref{subsection:colors} for a discussion.}
\vspace{0.1in}
\end{center}

\subsection{Notes on the Host Galaxies of Individual Sources}
\label{subsection:notes}

{\bf PKS\,0138--097:} The host galaxy of this high redshift ($z=0.733$)  
BL Lac object remains unresolved despite several attempts to separate the
two components \citep[e.g.][]{Hei96,Sca99a,Sca00a}.  We masked out its
three brightest companions before extracting the radial profile, but
included the region in which the faintest companion (D) is embedded since
it was found to contribute $\ll 1 \%$ of the total flux at K'-band
\citep{Hei96}. We found that our K-band profile is consistent with a 14.6
mag PSF, very near the K' value of 14.4 mag found by \citet{Hei96}.  
These values are nearly two magnitudes fainter than this object's
historical K-band peak \citep{Fan99}.

{\bf MS\,0158+002:} The host galaxy was clearly resolved by \hst\
\citep{Sca00a,Urr00} and ground-based observatories \citep{Wur96,Fal99} as
an elliptical with $M_{R} = -23.0$ mag, $\mu_{e} = 21.9$ mag
arcsec$^{-2}$, and $r_{e} = 9.3$ kpc (\hst\ values).  There is a close
companion to the N-W which we masked out in our analysis of the radial
profile. We clearly detected the host (the first time in a NIR band),
which appears to be a typical elliptical galaxy.

{\bf MS\,0331--363:} An \hst\ snapshot \citep{Sca00a,Urr00} resolved the
host as a giant ($r_{e} = 18.7$ kpc) elliptical galaxy with $M_{R} =
-24.3$ mag and $\mu_{e} = 22.1$ mag arcsec$^{-2}$. We have made the first
detection of the host galaxy in the NIR and it is the largest galaxy
($r_{e} = 8.5$ kpc) of those we observed.
 
{\bf 1ES\,0347--121:} \citet{Fal00} found a very circular elliptical
galaxy hosting this BL Lac object from a 2-D structural analysis of the
\hst\ survey data \citep{Sca00a,Urr00}.  The \hst\ data gave best fits of
$M_{R} = -23.2$ mag and $\mu_{e} = 20.6$ mag arcsec$^{-2}$ which are
consistent with ground-based values \citep{Wur96,Fal99}. However, the
half-light radius obtained by \hst\ ($r_{e} = 5.3$ kpc) is about twice
larger than the ground value \citep{Wur96,Fal99}.  We resolve the host
galaxy clearly, which is the first NIR detection of the host. Our
value of $r_{e} = 4.6$ kpc is slightly smaller than the \hst\ value but is
consistent with the observation by \citet[their Figure~6]{Sca00b} that BL
Lac hosts tend to be bluer towards their outer regions.

{\bf MS\,0350--371:} This BL Lac's host galaxy is a normal elliptical as
determined from 2-D analysis of the \hst\ image by \citet{Fal00}. The
\hst\ data are best fit by an elliptical galaxy with $M_{R} = -23.4$ mag,
$\mu_{e} = 20.8$ mag arcsec$^{-2}$, and $r_{e} = 6.5$ kpc
\citep{Sca00a,Urr00}. We detected the host galaxy in two NIR bands --
the first detections of the host at these wavelengths -- and derived a
NIR color ($J-K = 0.7$ mag) very near the average in our sample.
            
{\bf MS\,0419+194:} The host has been resolved in the optical band as an
elliptical galaxy \citep{Wur96,Fal99,Sca00a,Urr00}.  However, at
$z=0.514$, the host galaxy was unresolved in our mildly deep observation
(27 min). Our analysis show that the 1-D K-band profile is consistent with
a 16$^{th}$ magnitude point source, very near the 15.4 mag observed by
\citet{Gea93} during 1992.

{\bf PKS\,0521--365:} This object contains a well-studied optical/radio
jet extending to the N-W \citep[see][and references therein]{Sca99b} which
was recently detected in the X-rays \citep{Bir02}. Its host was resolved
in early ground-based imaging \citep{Fal94,Wur96}, which are consistent
with subsequent \hst\ results \citep[$M_{R} = -23.3$ mag, $\mu_{e} = 19.9$
mag arcsec$^{-2}$, and $r_{e} = 4.1$ kpc;][]{Sca00a,Urr00}.  H-band
imaging from the ground \citep{Kot98b} shows an intermediate galaxy
absolute magnitude to our J and K results ($M_{H} = -25.4$ mag), but for a
larger galaxy ($r_{e}=3.8$ kpc, versus our range of 1.5 to 2.5 kpc).  
\citet{Fal00} found this to be an average elliptical galaxy from a 2-D
analysis of the \hst\ data.

{\bf H\,1914--194:} This is a little studied BL Lac object with no
redshift determined. \hst\ results show $m_{R} = 17.0$ mag, $\mu_{e} =
24.7$ mag arcsec$^{-2}$, and $r_{e} = 7\arcsec.4$ for an elliptical host,
but this was a rare instance in which a disk profile fits the \hst\ data
equally well \citep{Sca00a,Urr00}. We fit a slightly smaller host ($r_{e}
= 4\arcsec$) in both of our K-band images, which are the first detections
of the galaxy in the NIR.
            
{\bf PKS\,2005--489:} The large round elliptical host of this well-known
blazar has been much studied. Our H-band results are remarkably consistent
(within 1$\sigma$) with those obtained by \citet{Kot98b}. We resolved the
bright galaxy in all three NIR bands even with short four to twelve minute
exposures. Best fits from \hst\ data \citep[$M_{R} = -23.9$ mag, $\mu_{e}
= 21.3$ mag arcsec$^{-2}$, and $r_{e} = 10.5$ kpc;][]{Urr00,Sca00a,Fal00}
show a bright elliptical host with derived parameters that agree with
previous ground-based imaging \citep{Sti93,Fal96}.  However, the effective
radii determined by the optical imaging is consistently larger than values
derived in the NIR (\citet{Kot98b}; this work) -- up to three times larger
than the J-band value, and five times larger at K.  Our observations of
this blazar were coordinated near a multi-wavelength monitoring campaign
\citep{Per99,Tag01} in which both short and long-term X-ray outbursts were
observed. We found no significant variability of the nucleus between the
nights in our NIR bands.
            
{\bf MS\,2143+070:} The host galaxy of this BL Lac object has been well
studied with \hst\ and ground-based observatories.  \citet{Sca00b} derived
$m_{H} = 15.3$ mag from NICMOS data, which is consistent with the
ground-based value derived by \citet{Kot98b}. These H-band values are
intermediate to our derived J and K galaxy magnitudes, as expected, and
the effective radii we determined (6.0 and 7.1 kpc, at K and J,
respectively) are also within the range of H-band derived values from the
ground (5.5 kpc) and space (7.6 kpc). An I-band \hst\ detection of the
host by \citet{Jann97} was confirmed \citep{Urr99} and combined with
V-band data from the \hst\ survey \citep{Sca00a,Urr00} and found to have a
$V-I$ color of 1.61 mag. The \hst\ survey results show $M_{R} = -23.7$
mag, $\mu_{e} = 21.7$ mag arcsec$^{-2}$, and $r_{e} = 10.6$ kpc
\citep{Sca00a,Urr00}, assuming $V-R = 0.86$ mag, which agree well with
ground-based results \citep{Wur96,Fal99}.

{\bf PKS\,2240--260:} The redshift of 0.774 for this object is based on
two weak features in the spectra and awaits confirmation \citep{Sti93}.  
The host was unresolved in the \hst\ snapshot survey \citep{Sca00a,Urr00}
and remains unresolved by us, even with our fairly deep (42 min) image.
This is the first published attempt to resolve the host in the NIR. We
find a K-band magnitude for the BL Lac of 13.3 mag that is near its
historical maximum in this band \citep{Fan99}. It has been seen to be as
faint as 14.8 mag at this band \citep{Gea93}.

{\bf H\,2356--309:} The host is resolved as an elliptical in the \hst\
data with $M_{R} = -23.2$ mag, $\mu_{e} = 21.1$ mag arcsec$^{-2}$, and
$r_{e} = 7.1$ kpc \citep{Sca00a,Urr00} in general agreement with previous
ground-based results \citep{Fal91}. \citet{Fal00} found that this is a
normal elliptical from a 2-D analysis of the \hst\ data and also found a
close 1$\arcsec$.2 companion to the S-W. This companion is
indistinguishable in our images and can contribute only $<1\%$ to the
total flux of the galaxy \citep{Fal00} so was not masked. We imaged the
galaxy in three NIR bands and our two epochs of H-band results agree well
within our errors, and with results from NICMOS imaging \citep[$M_{H} =
-25.5$ mag and $\mu_{e} = 18.5$ mag arcsec$^{-2}$;][]{Sca00b} except that
they fit a larger galaxy ($r_{e} = 6.5$ kpc versus our derived value of
4.2 kpc) with smaller uncertainty.

\section{Summary and Conclusions}
\label{section:conclusions}

We have successfully separated the NIR host galaxy emission from the
bright nuclei of eight BL Lac objects. These objects are at known, and
relatively low ($z \leq 0.308$) redshifts.  Three targets at $z \simgt
0.5$ were unresolved, and one marginally resolved BL Lac is at an unknown
redshift. The hosts studied here are large elliptical galaxies with
average luminosities comparable to brightest cluster member galaxies and
over one magnitude brighter than normal ellipsoids (see
section~\ref{subsection:morphology}). This is consistent with findings
established in previous H-band \citep{Kot98b,Sca00b} and optical waveband
studies \citep[e.g.][]{Fal96,Wur96,Sca00a,Urr00}. Our additional data
allowed us to better constrain a Kormendy relation for BL Lac objects in
the NIR which is remarkably similar to that of normal ellipticals and in
general agreement with observations of radio galaxies and other types of
AGN \citep[e.g.][see section~\ref{subsection:morphology}]{Tay96}.

Our observations allowed us to investigate colors within the NIR-band for
the first time in five galaxies, as well as providing additional
optical-to-NIR colors for eight objects. We find average $R-K$ and $J-K$
colors which are indistinguishable from those measured in normal
elliptical galaxies (section~\ref{subsection:colors}), typical of old
stellar populations at the redshifts of the sources \citep{Pog97,Sca00b}.
The color profiles which utilized R-band \hst\ observations show a bluer
trend away from the galaxy centers, although the magnitude of this
gradient is not definitive. Taken together, the presently available NIR
data show that in all respects we could investigate, the BL Lac host
galaxies are luminous ellipticals showing old stellar populations, and
lends further support to a picture in which all elliptical galaxies can
experience a phase of nuclear activity and that the galaxies do not know
about the nucleus. Considering recent results on black hole demography
\citep[see][for a recent summary]{Urr03}, which shows that most probably
all ellipticals harbor a super-massive black hole in their center
\citep[e.g.][]{Geb00,Fer01}, such a finding is not very surprising.


\acknowledgments

Support for this work was provided by NASA through grant number
GO-07893.01-96A from the Space Telescope Science Institute, which is
operated by AURA, Inc., under NASA contract NAS~5-26555. C.~C.~C. was a
STScI Summer Student Program participant in 2000, and is thankful to its
coordinator David Soderblom, and is also grateful to John Wardle and Dave
Roberts at Brandeis for their continued support of his research. C.~M.~U.  
and M.~G. thank the staff of the Las Campanas Observatory for their
hospitality and support during the observing run. This research has made
use of NASA's Astrophysics Data System Abstract Service.



\newpage

\begin{deluxetable}{lcccccc}
\label{tbl-1}
\tablecolumns{7} 
\tablewidth{0pc} 
\tablecaption{Journal of the Observations\label{tbl-1}}
\tablehead{ 
\colhead{Name}                  &\colhead{Date}
&\colhead{Filter}               &\colhead{Exposure*}
&\colhead{Sky Brightness}       &\colhead{FWHM}
&\colhead{Resolved?}
\\ 
\colhead{}                      &\colhead{(in 1998)} 
&\colhead{}                     &\colhead{[minutes]}
&\colhead{[mag arcsec$^{-2}$]}  &\colhead{[arcsec]} 
&\colhead{}
}  
\startdata 
PKS\,0138--097 &Sep 29  &K              &54     &13.4   &0.79 &No\\
MS\,0158+002   &Sep 30  &K              &72     &13.4   &0.93 &Yes\\
MS\,0331--363  &Sep 29  &K              &36     &13.4   &0.83 &Yes\\
               &Oct 1   &K              &18     &13.5   &1.05 &Yes\\
1ES\,0347--121 &Sep 30  &K              &54     &13.3   &0.87 &Yes\\
MS\,0350--371  &Sep 30  &K              &18     &13.2   &0.93 &Yes\\
               &Oct 1   &K, J           &36, 36 &13.5, 16.2&0.92, 1.10 
&Yes\\
MS\,0419+194   &Sep 28  &K              &27     &13.5   &0.93 &No\\
PKS\,0521--365 &Oct 1   &K, J           &28, 18 &13.3, 15.8&1.01, 
0.91&Yes\\
H\,1914--194   &Sep 28  &K              &63     &13.4   &0.95 &Yes\\
               &Sep 30  &K              &54     &13.2   &0.72 &Yes\\
PKS\,2005--489 &Sep 29  &K, J, H&4.5, 12, 4.5&13.2, 15.8, 13.6&0.78, 0.89, 
0.81 &Yes\\
               &Oct 1   &K, J, H        &4, 12, 7.5&13.2, 15.4, 13.6&0.93, 
0.93, 0.88 &Yes\\
MS\,2143+070   &Sep 29  &K              &54     &13.2   &0.89 &Yes\\
               &Sep 30  &J              &54     &15.8   &0.92 &Yes\\
PKS\,2240--260 &Oct 1   &K              &42     &13.4   &0.93 &No\\
H\,2356--309   &Sep 28  &K, J, H&54, 54, 27&13.8, 16.7, 14.3&0.82,0.88, 
0.86 &Yes\\
               &Sep 29  &H              &54     &14.1   &0.86 &Yes\\
\enddata 
\tablecomments{*Total integration time for each night.
}
\end{deluxetable}

\begin{deluxetable}{lcccccccccccc}
\label{tbl-2}
\rotate
\tablecolumns{13} 
\tablewidth{0pc} 
\tablecaption{BL Lac and Host Galaxy Near-Infrared Properties\label{tbl-2}}
\tablehead{ 
\colhead{Name}                  &\colhead{$z$} 
&\colhead{Filter}               &\colhead{$A^{(a)}$}
&\colhead{$K$-corr$^{(b)}$}     &\colhead{$m_{Tot}^{(c)}$} 
&\colhead{$m_{PSF}^{(c)}$}      &\colhead{$m_{Host}^{(c)}$} 
&\colhead{$\mu_{e}^{(d)}$}      &\colhead{$r_{e}^{(e)}$}
&\colhead{$M_{PSF}^{(f)}$}      &\colhead{$M_{Host}^{(f)}$} 
&\colhead{$r_{e}^{(e)}$}
\\ 
\colhead{}                      &\colhead{} 
&\colhead{}                     &\colhead{[mag]}
&\colhead{[mag]}                &\colhead{[mag]} 
&\colhead{[mag]}                &\colhead{[mag]} 
&\colhead{[mag/asec$^{2}$]}     &\colhead{[arcsec]}
&\colhead{[mag]}                &\colhead{[mag]} 
&\colhead{[kpc]}
}  
\startdata 
PKS 0138--097    &0.733 &K       &0.02 &...      &14.6  &...             
&...             &...               &...                &--29.3  &...    
&...            \\
MS  0158+002    &0.299 &K       &0.02 &--0.65    &14.9  &15.9 $\pm$ 0.6  
&14.6 $\pm$ 0.5  &17.0 $\pm$ 3.6      &0.8 $\pm$ 1.3    &--25.7  &--26.3  
&4.7 $\pm$ 7.7  \\
MS  0331--363   &0.308 &K       &0.01 &--0.65    &14.7  &16.3 $\pm$ 0.8  
&14.5 $\pm$ 0.3  &18.1 $\pm$ 2.7      &1.4 $\pm$ 1.7    &--25.4  &--26.5  
&8.5 $\pm$ 10.3   \\
1ES 0347--121   &0.185 &K       &0.04 &--0.48    &14.5  &15.8 $\pm$ 0.5  
&14.2 $\pm$ 0.3  &16.2 $\pm$ 1.1      &0.6 $\pm$ 0.3    &--24.7  &--25.8  
&2.5 $\pm$ 1.3    \\
MS  0350--371   &0.165 &K       &0.01 &--0.44    &13.9  &15.4 $\pm$ 0.5  
&13.7 $\pm$ 0.2  &17.3 $\pm$ 1.5      &1.2 $\pm$ 0.8    &--24.8  &--26.1 
&4.6 $\pm$ 3.1    \\
                &      &J       &0.03 &0.09     &15.3  &17.2 $\pm$ 1.1  
&14.9 $\pm$ 0.4  &17.5 $\pm$ 2.6      &1.0 $\pm$ 1.2    &--23.0  &--25.4  
&3.8 $\pm$ 4.6    \\
MS  0419+194    &0.512 &K       &0.18 &...      &16.0  &...             
&...             &...               &...                &--27.1  &...    
&...            \\
PKS 0521--365   &0.055 &K       &0.03 &--0.18    &11.3  &12.4 $\pm$ 0.3  
&11.4 $\pm$ 0.2  &14.7 $\pm$ 0.5      &1.0 $\pm$ 0.2    &--25.3  &--26.1  
&1.5 $\pm$ 0.3    \\
                &      &J       &0.07 &0.03     &12.7  &14.1 $\pm$ 0.1  
&12.7 $\pm$ 0.1  &16.9 $\pm$ 0.3      &1.7 $\pm$ 0.2    &--23.6  &--25.1  
&2.5 $\pm$ 0.3  \\ 
H   1914--194   &...   &K       &...  &...      &12.7  &12.8 $\pm$ 0.1  
&14.2 $\pm$ 0.4  &...                 &4.0 $\pm$ 1.9    &...    &...    
&...            \\
PKS 2005--489   &0.071 &K       &0.01 &--0.22    &10.2  &10.6 $\pm$ 0.2  
&11.1 $\pm$ 0.2  &14.4 $\pm$ 0.3      &1.0 $\pm$ 0.1    &--27.6  &--26.9  
&1.9 $\pm$ 0.2    \\
                &      &H       &0.02 &0.00     &10.7  &11.1 $\pm$ 0.1  
&11.7 $\pm$ 0.2  &15.7 $\pm$ 0.5      &1.5 $\pm$ 0.3    &--27.1  &--26.5  
&2.8 $\pm$ 0.6    \\
                &      &J       &0.03 &0.04     &11.3  &11.7 $\pm$ 0.1  
&12.6 $\pm$ 0.2  &17.1 $\pm$ 0.4      &2.0 $\pm$ 0.3    &--26.6  &--25.7  
&3.7 $\pm$ 0.6    \\
MS  2143+070    &0.237 &K       &0.04 &--0.57    &14.4  &15.7 $\pm$ 0.5  
&14.3 $\pm$ 0.3  &17.7 $\pm$ 2.6      &1.2 $\pm$ 1.4    &--25.4  &--26.2  
&6.0 $\pm$ 7.1    \\
                &      &J       &0.11 &0.11     &15.9  &16.8 $\pm$ 0.5  
&15.8 $\pm$ 0.5  &18.8 $\pm$ 2.2      &1.4 $\pm$ 1.4    &--24.3  &--25.4  
&7.1 $\pm$ 7.1    \\
PKS 2240--260   &0.774$^{(g)}$&K       &0.01 &...      &13.3  &...             
&...             &...               &...                &--30.8  &...    
&...            \\
H   2356--309   &0.165 &K       &0.01 &--0.44    &13.8  &14.5 $\pm$ 0.2  
&14.1 $\pm$ 0.2  &17.5 $\pm$ 1.6      &1.1 $\pm$ 0.8    &--25.7  &--25.6  
&4.2 $\pm$ 3.1    \\
                &      &H       &0.02 &0.01     &14.3  &15.1 $\pm$ 0.3  
&14.5 $\pm$ 0.3  &17.4 $\pm$ 2.2      &1.1 $\pm$ 1.1    &--25.1  &--25.7  
&4.2 $\pm$ 4.2    \\
                &      &J       &0.03 &0.09     &15.1  &15.9 $\pm$ 0.4  
&15.2 $\pm$ 0.4  &18.0 $\pm$ 1.6      &1.1 $\pm$ 1.7    &--24.3  &--25.1  
&4.2 $\pm$ 6.5  \\

\hline
K-band median$^{(h)}$&0.175 &        &     &         
&&15.6&14.2&&&--25.4&--26.2&4.4\\
\enddata 
\tablecomments{$^{(a)}$ Interstellar extinction corrections using R-band 
extinction coefficients ($A_{R}$) from \citet{Urr00} and appropriate NIR 
relations from \citet{Car89}: $A_{J}/A_{R}=0.377$, $A_{H}/A_{R}=0.234$, 
and $A_{K}/A_{R}=0.150$. \\ 
$^{(b)}$ K-corrections for first-ranked elliptical galaxies from \citet{Neu85} 
citing private communication from S.E. Persson (1984), interpolated to
the redshifts of our targets. \\
$^{(c)}$ Observed total (Tot), BL Lac nucleus (PSF), and (Host) galaxy 
magnitudes. The host galaxy magnitudes were integrated to infinite radius 
so will appear in many cases to exceed the observed total magnitudes.\\
$^{(d)}$ Surface brightness at the half-light (effective) radius 
($r_{e}$) with corrections for extinction, K-correction, and cosmological dimming applied. \\
$^{(e)}$ Bulge scale-length (effective radius), i.e., the radius at which 
half of the light in the galaxy is enclosed. \\
$^{(f)}$ Absolute magnitudes for the BL Lac nucleus (PSF) and Host. Both
were extinction corrected but only the galaxy values were K-corrected. The
BL Lac nuclei are assumed to have flat power law spectra, therefore have
negligible K-corrections. \\
$^{(g)}$ Uncertain redshift. \\
$^{(h)}$ Statistics only for resolved objects. \\
}
\end{deluxetable}

\begin{deluxetable}{lccc}
\label{tbl-3}
\tablecolumns{4} 
\tablewidth{0pc} 
\tablecaption{Host Galaxy Colors \label{tbl-3}}
\tablehead{ 
\colhead{Name}                  &\colhead{R--K$^{(a)}$}
&\colhead{J--K$^{(a)}$}         &\colhead{$\Delta(R-K)/\Delta log~r~^{(b)}$} 
\\ 
\colhead{}                      &\colhead{[mag]} 
&\colhead{[mag]}                &\colhead{[mag]} 
}  
\startdata 
MS\,0158+002   &3.3     & ... & --2.00 \\
MS\,0331--363  &2.2     & ... & --1.10 \\
1ES\,0347--121 &2.6     & ... & --1.06 \\
MS\,0350--371  &2.7     & 0.7 & --0.86 \\
PKS\,0521--365 &2.8     & 1.0 & --1.86 \\
PKS\,2005--489 &3.0     & 1.2 & --2.37 \\
MS\,2143+070   &2.5     & 0.8 & --0.33 \\
H\,2356--309   &2.4     & 0.5 & --0.57 \\
\hline
Median         &2.7     & 0.8 & --1.08 \\
\enddata 
\tablecomments{$^{(a)}$ Integrated rest frame colors, extinction and
K-corrected. R-band values are taken from \citet{Urr00}. \\
$^{(b)}$ Color gradient measured from the data presented in 
Figure~\ref{fig-4}
which used our K-band results and the R-band data from the \hst\ survey 
\citep{Sca00a,Urr00}.\\
}
\end{deluxetable}

\end{document}